\documentclass{ws-procs9x6}

\begin{document}

\title{Locking Internal and Space-Symmetries: \\Relativistic Vector
Condensation}

\author{Francesco SANNINO\footnote{\uppercase{T}his work is supported by
the \uppercase{M}arie--\uppercase{C}urie fellowship under contract
\uppercase{MCFI}-2001-00181.}}
\address{{\rm NORDITA}, Blegdamsvej 17, DK-2100 Copenhagen \O, Denmark \\
E-mail: francesco.sannino@nbi.dk}


\maketitle

\abstracts{Internal and Lorentz symmetries are necessarily linked
 when considering non scalar condensates.
Here I review vectorial type condensation due to a non zero
chemical potential associated to some of the global conserved
charges of the theory. The phase structure is very rich since
three distinct phases exists depending on the value assumed by one
of the zero chemical potential vector self interaction terms. In a
certain limit of the couplings and for large chemical potential
the theory is not stable. This limit corresponds to a gauge type
limit often employed to economically describe the ordinary vector
mesons self interactions in QCD. Our analysis is relevant since it
leads to a number of physical applications not limited to strongly
interacting theories at non zero chemical potential.}

\section{Introduction}
\label{introduction}

Relativistic vector condensation has been proposed and studied in
different realms of theoretical physics. However the condensation
mechanism and the nature of the relativistic vector mesons
themselves is quite different. Linde \cite{Linde:1979pr}, for
example, proposed the condensation of the intermediate vector
boson $W$ in the presence of a superdense fermionic matter while
Ambj\o rn and Olesen \cite{Ambjorn:1989gb,{Kajantie:1998rz}}
investigated their condensation in presence of a high external
magnetic field. Manton \cite{Manton:1979kb} and later on Hosotani
\cite{Hosotani:1988bm} considered the extension of gauge theories
in extra dimensions and suggested that when the extra dimensions
are non simply connected the gauge fields might condense. Li in
\cite{Li:2002iw} has also explored a simple effective Lagrangian
and the effects of vector condensation when the vectors live in
extra space dimensions\footnote{The effective Lagrangian and the
condensation phenomenon in extra (non compact) space dimensions
has also been studied/suggested in \cite{Sannino:2001fd}.}.

In \cite{Brown:kk,{Langfeld}} it has also been suggested that the
non gauge vectors fields such as the (quark) composite field
$\rho$ in QCD may become light and possibly condense in a high
quark matter density and/or in hot QCD. Harada and Yamawaki's
\cite{Harada:2000kb} dynamical computations within the framework
of the hidden local gauge \cite{BKY} symmetry support this
picture.

We consider another type of condensation. If vectors themselves
carry some global charges we can introduce a non zero chemical
potential associated to some of these charges. If the chemical
potential is sufficiently high one can show that the gaps (i.e.
the energy at zero momentum) of these vectors become light
\cite{{Sannino:2002wp},{Sannino:2001fd},{Lenaghan:2001sd}} and
eventually zero signaling an instability. If one applies our
results to 2 color Quantum Chromo Dynamics (QCD) at non zero
baryon chemical potential one predicts that the vectors made out
of two quarks (in the 2 color theory the baryonic degrees of
freedom are bosons) condense. Recently lattice studies for 2 color
at high baryonic potential \cite{Alles:2002st} seem to support our
predictions. This is the relativistic vectorial Bose-Einstein
condensation phenomenon. A decrease in the gap of vectors is also
suggested at high baryon chemical potential for two colors in
\cite{Muroya:2002ry}.

I review here the general structure of vector condensation
\cite{Sannino:2002wp}.

\section{Vacuum Structure and Different Phases}
\label{vacuum} We choose to consider the following general
effective Lagrangian for a relativistic massive vector field in
the adjoint of $SU(2)$ in $3+1$ dimensions and up to four vector
fields, two derivatives and containing only intrinsic positive
parity terms \cite{ARS,{DRS}}\footnote{{}For simplicity and in
view of the possible physical applications we take the vectors to
belong to the adjoint representation of the $SU(2)$ group.}:
\begin{eqnarray}
{
L}&=&-\frac{1}{4}F^a_{\mu\nu}F^{a{\mu\nu}}+\frac{m^2}{2}A_{\mu}^a
A^{a\mu} +\delta\, \epsilon^{abc}\partial_{\mu} A_{a\nu} A^{\mu}_b
A^{\nu}_c \nonumber \\ &-&
\frac{\lambda}{4}\left(A^a_{\mu}A^{a{\mu}}\right)^2 +
 \frac{\lambda^{\prime}}{4} \left(A^a_{\mu}A^{a{\nu}}\right)^2 \ ,
 \end{eqnarray}
with $F_{\mu
\nu}^a=\partial_{\mu}A^a_{\nu}-\partial_{\nu}A^a_{\mu}$, $a=1,2,3$
and metric convention $\eta^{\mu \nu}={\rm diag}(+,-,-,-)$. Here,
$\delta$ is a real dimensionless coefficient, $m^2$ is the tree
level mass term and $\lambda$ and $\lambda^{\prime}$ are positive
dimensionless coefficients with $\lambda \ge \lambda^{\prime}$
when $\lambda^{\prime}\ge 0$ or $\lambda\ge 0$ when
$\lambda^{\prime}\ge 0$ to insure positivity of the potential. The
Lagrangian describes a self interacting $SU(2)$ Yang-Mills theory
in the limit $m^2=0$, $\lambda=\lambda^{\prime}>0$ and $
\delta=-\sqrt{\lambda}$.

We set $\delta=0$ and the theory gains a new symmetry according to
which we have always a total number of even vectors in any process
\cite{Sannino:2002wp}. The effect of a nonzero chemical potential
associated to a given conserved charge - (say $T^3=\tau^3/2$) -
can be included by modifying the derivatives acting on the vector
fields according to $\partial_{\nu} A_{\rho} \rightarrow
\partial_{\nu}A_{\rho} - i \left[B_{\nu}\ ,A_{\rho}\right]$
with $B_{\nu}=\mu \,\delta_{\nu 0} T^3\equiv V_{\nu} T^3$ where
$V=(\mu\ ,\vec{0})$. The chemical potential breaks explicitly the
Lorentz transformation leaving invariant the rotational symmetry.
Also the $SU(2)$ internal symmetry breaks to a $U(1)$ symmetry. If
the $\delta$ term is absent we have an extra unbroken $Z_2$
symmetry which acts according to $A^3_{\mu}\rightarrow
-A^3_{\mu}$. These symmetries suggest introducing the following
cylindric coordinates:
\begin{eqnarray}
\phi_\mu &=& {1 \over \sqrt{2}} ( A^1_\mu + i A^2_\mu )\ , \qquad
\phi^{\ast}_\mu = {1 \over \sqrt{2}} ( A^1_\mu - i A^2_\mu )\ ,
\qquad \psi_\mu = A^3_\mu \ ,
\end{eqnarray}
on which the covariant derivative acts as follows:
\begin{eqnarray}
D_{\mu}\phi_{\nu} = \left(\partial + i\, V \right)_{\mu}
\phi_{\nu} \ , \quad  D_{\mu}\psi_{\nu} = \partial_{\mu}
\psi_{\nu} \ , \quad V_\nu = \left(\mu,{\bf 0}\right) \ .
\end{eqnarray}

The vacuum structure of the theory is explored via the variational
ansatz \cite{Sannino:2002wp}:
\begin{eqnarray}
\psi^{\mu}= 0\ , \qquad \phi^{\mu} = \sigma \,
\left(%
\begin{array}{c}
  0 \\
  1 \\
    e^{i\alpha}\\
  0 \\
\end{array}%
\right) \ .
\end{eqnarray}
Substituting the ansatz in the potential expression yields:
\begin{eqnarray}
V=2\, \sigma^4\left[(2\lambda - \lambda^{\prime})-
\lambda^{\prime}\cos^2\alpha \right] + 2\left(m^2 - \mu^2\right)
\sigma^2 \ .
\end{eqnarray}
The potential is positive for any value of $\alpha$ when $\lambda
> \lambda^{\prime}$ if $\lambda^{\prime}\ge 0$ or $\lambda >0$ if
$\lambda^{\prime}<0$. Due to our ansatz the ground state is
independent of $\delta$. The unbroken phase occurs when $\mu \leq
m$ and the minimum is at $\sigma=0$. A possible broken phase is
achieved when $\mu> m$ since in this case the quadratic term in
$\sigma$ is negative. According to the value of $\lambda^{\prime}$
we distinguish three distinct phases:
\subsection{The polar phase: $\lambda^{\prime}>0$}
In this phase the minimum is for
\begin{eqnarray}
 \langle \phi^{\mu} \rangle = \sigma \,
\left(%
\begin{array}{c}
  0 \\
  1 \\
  1\\
  0 \\
\end{array}%
\right) \ , \quad {\rm with} \quad \sigma^2=\frac{1}{4}\frac{\mu^2
- m^2}{\lambda - \lambda^{\prime} }\ ,
\end{eqnarray}
and we have the following pattern of symmetry breaking
$SO(3)\times U(1) \rightarrow SO(2)$, where $SO(3)$ is the
rotational group. We have three broken generators and 3 gapless
excitations with linear dispersion relations. All of the physical
states (with and without a gap) are either vectors (2-component)
or scalars with respect to the unbroken $SO(2)$ group. The
dispersion relations for the 3 gapless states can be found in
\cite{Sannino:2002wp}.

At $\mu=m$ the dispersion relations are no longer linear in the
momentum. This is related to the fact that the specific part of
the potential term has a partial conformal symmetry discussed
first in \cite{Sannino:2001fd}. Some states in the theory are
curvatureless but the chemical potential present in the derivative
term prevents these states to be gapless. There is a transfer of
the conformal symmetry information from the potential term to the
vanishing of the velocity of the gapless excitations related to
the would be gapless states. This conversion is due to the linear
time-derivative term induced by the presence of the chemical
potential term
\cite{Schafer:2001bq,{Sannino:2001fd},{Sannino:2002wp}}.

\subsection{Enhanced symmetry and type II Goldstone bosons: $\lambda^{\prime}=0$}
Here the potential has an enhanced $SO(6)$ in contrast to the
$SU(2)\times U(1)$ for $\lambda^{\prime}\neq 0$ global symmetry
which breaks to an $SO(5)$ with 5 broken generators. Expanding the
potential around the vacuum we find 5 null curvatures
\cite{Sannino:2002wp}. However we have only three gapless states
obtained diagonalizing the quadratic kinetic term and the
potential term. Two states (a vector of $SO(2)$) become type II
goldstone bosons while the scalar state remains type I
\cite{Nielsen}. This latter state is the goldstone boson related
to the spontaneously broken $U(1)$ symmetry\footnote{According to
the Nielsen and Chadha counting scheme in absence of Lorentz
invariance if $n_I$ denotes the number of gapless excitations of
type $I$ with linear dispersion relations (i.e. $E\propto p$) and
$n_{II}$ the ones with quadratic dispersion relations (i.e.
$E\propto p^2$) the Goldstone theorem generalizes to
$n_I+2\,n_{II} \ge \#~{\rm broken~generators}$.}. Using the
Nielsen Chadha theorem\cite{Nielsen} the type II states are
counted twice with respect to the number of broken generators
while the linear just once recovering the number of generators
broken by the vacuum. We can be more specific since we discovered
\cite{Sannino:2001fd} that the velocity of the associated gapless
states is proportional to the curvatures (evaluate on the minimum)
of the would be goldstone bosons which is zero in the
$\lambda^{\prime}=0$ limit. Again we have an efficient mechanism
for communicating the information of the extra broken symmetries
from the curvatures to the velocities of the already gapless
excitations.
\subsection{The apolar phase: $\lambda' < 0$}
In this case the potential is minimized for:
\begin{eqnarray}
\langle \phi^{\mu} \rangle= \sigma \,
\left(%
\begin{array}{c}
  0 \\
  1 \\
  \imath\\
  0 \\
\end{array}%
\right) \ , \quad {\rm with} \quad\sigma^2 =\frac{1}{2}\frac{\mu^2
- m^2}{2\lambda - \lambda^{\prime}} \ ,
\end{eqnarray}
with 3 broken generators. However the unbroken generator is a
linear combination of the $U(1)$ and a $SO(3)$ generator and in
\cite{Sannino:2002wp} it has been shown that only two gapless
states emerges. One of the two states is a type I goldstone boson
while the other is type II. The two goldstone bosons are one in
the $z$ and the other in the $x-y$ plane. Interestingly in this
phase, due to the intrinsic complex nature of the vev, we have
spontaneous $CP$ breaking. We summarize in Fig.~1 the phase
structure in terms of the number of goldstone bosons and their
type according to the values assumed by $\lambda^{\prime}$.
\begin{figure}[h]\vskip -.2cm
\begin{center}\includegraphics[width=5truecm,
height=3truecm]{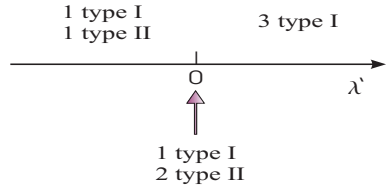} \vskip -.7cm \caption{We show the
number and type of goldstone bosons in the three distinct phases
associated to the value assumed by the coupling
$\lambda^{\prime}$. In the polar phase, positive
$\lambda^{\prime}$, we have 3 type I goldstone bosons. In the
apolar phase, negative $\lambda^{\prime}$, we have one type I and
one type II goldstone boson while in the enhanced symmetry case
$\lambda^{\prime}=0$ we have one type I and two type II
excitations.} \end{center} \label{Figura1}
\end{figure}
\subsection{The case $\lambda = \lambda^{\prime}$: the gauge theory
limit} 
\begin{figure}[hbt]\begin{center}
\includegraphics[width=6truecm, height=4cm]{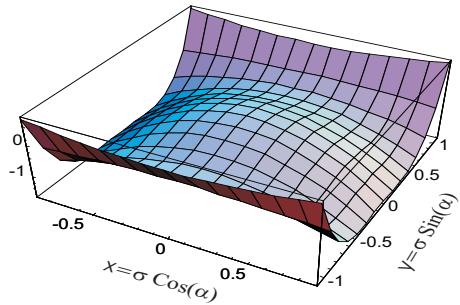}
\caption{Potential plotted for $\mu=2m$ and
$\lambda=\lambda^{\prime}=1$.} \label{gauge} \end{center}
\end{figure}

Here the potential is:
\begin{eqnarray}
V=2\, \sigma^4\lambda \sin^2\alpha  + 2\left(m^2 - \mu^2\right)
\sigma^2 \ ,
\end{eqnarray}
with two extrema when $\mu>m$, one for $\alpha=0$ and $\sigma=0$
which is an unstable point and the other for $\alpha=\pm\pi/2$ and
$\sigma^2=\frac{(\mu^2-m^2)}{2\lambda}$ corresponding to a saddle
point (see the potential in Fig.~\ref{gauge}).
At first the fact that we have no stable solutions seems
unreasonable since we know that in literature we often encounter
condensation of intermediate vector mesons such as the $W$ boson.
However (except for extending the theory in higher space
dimensions) in these cases one often introduces an external
source. {}For example one adds to the theory a strong magnetic
field (say in the direction $z$) which couples to the
electromagnetically charged intermediate vector bosons $W^{+}$ and
$W^{-}$). In this case the potential is (see
\cite{Ambjorn:1989gb}):
\begin{eqnarray}
V=2\, \sigma^4\lambda \sin^2\alpha  + 2\left(m^2 -
e\,H\,\sin\alpha\right) \sigma^2 \ ,
\end{eqnarray}
where $e$ is the electromagnetic charge and $H$ is the external
electromagnetic source field. This potential has a true minimum
for $\alpha=\pi/2$ and $\sigma=\frac{eH-m^2}{2\lambda}$ whenever
the external magnetic field satisfies the relation $eH>m$. We
learn that the relativistic vector theory is unstable at large
chemical potential whenever the non derivative vector self
interactions are tuned to be identical. This is precisely the
limit often used in literature when writing effective Lagrangians
that in QCD describe the $\rho$ vector field. In principle we can
still imagine to stabilize the potential in the gauge limit by
adding some higher order operators. This might be the case if one
introduces massive gauge bosons as in \cite{BKY}. Due to the gauge
limit $\lambda^{\prime}=\lambda$ is positive and assuming the
higher order corrections to be small one predicts a polar phase
within this model. Another solution to this instability is that
the chemical potential actually does not rise above the mass of
the vectors even if we increase the relative charge density. This
phenomenon is similar to what happens in the case of an ideal bose
gas at high chemical potential \cite{kapusta}. If the strict gauge
limit is taken (i.e. also the mass term is set to zero) we need to
impose gauge neutrality and the analysis modifies \cite{kapusta}.

Interestingly by studying the vector condensation phenomenon for
strongly interacting theories on the lattice at high isospin
chemical potential we can determine the best way of describing the
ordinary vector self-interactions at zero chemical potential.

\section{Physical Applications and Conclusions}
\label{conclusion}

We presented the phase structure of the relativistic massive
vector condensation phenomenon due to a non zero chemical
potential associated to some of the global conserved charges of
the theory \cite{Sannino:2002wp}. The phase structure is very
rich. According to the value assumed by $\lambda^{\prime}$ we have
three independent phases. The polar phase with $\lambda^{\prime}$
positive is characterized by a real vacuum expectation value and 3
goldstone bosons of type I. The apolar phase for
$\lambda^{\prime}$ negative has a complex vector vacuum
expectation value spontaneously breaking CP. In this phase we have
one goldstone boson of type I and one of type II while still
breaking 3 continuous symmetries. The third phase has an enhanced
potential type symmetry and 3 goldstone bosons one of type I and
two of type II.

We also discovered that if we force the self interaction couplings
$\lambda$ and $\lambda^{\prime}$ to be identical, as predicted in
a Yang-Mills massive theory, our ansatz for the vacuum does not
lead to a stable minimum when increasing the chemical potential
above the mass of the vectors. We suggest that lattice studies at
high isospin chemical potential in the vector channel for QCD
might be able to, indirectly, shed light on this sector of the
theory at zero chemical potential. More generally the hope is that
these studies might help understanding how to construct consistent
theories of interacting massive higher spin fields not necessarily
related to a gauge principle.

The present knowledge can be used for a number of physical
phenomena of topical interest. {}For example in the framework of 2
color QCD at non zero baryon chemical
potential\cite{Splittorff:2002xn} vector condensation has been
predicted in \cite{Lenaghan:2001sd,{Sannino:2001fd}}. Recent
lattice studies \cite{Alles:2002st} seem to support it.
Studies\cite{Hands:2003dh} of the Gross-Neveu model in 2+1
dimensions with a baryon chemical potential might also shed light
on the vector meson channel. The present analysis while
reinforcing the scenario of vector condensation shows that we can
have many different types of condensations with very distinct
signatures. {}Other possible physical applications are discussed
in \cite{Sannino:2001fd}. The analysis has been extended to a
general number of space dimensions \cite{Sannino:2001fd} and is
useful for various scenarios related to the phenomenon of vector
condensation \cite{Li:2002iw,{Moffat:2002nu}}.


\begin{thebibliography}{0}


\bibitem{Linde:1979pr}
A.~D.~Linde,
Phys.\ Lett.\ B {\bf 86}, 39 (1979).



\bibitem{Ambjorn:1989gb}
J.~Ambjorn and P.~Olesen,
Phys.\ Lett.\ B {\bf 218}, 67 (1989) [Erratum-ibid.\ B {\bf 220},
659 (1989)], ibid. \ B {\bf 257}, 201 (1991), Nucl.\ Phys.\ B {\bf
330}, 193 (1990).


\bibitem{Kajantie:1998rz}
K.~Kajantie, M.~Laine, J.~Peisa, K.~Rummukainen and
M.~E.~Shaposhnikov,
Nucl.\ Phys.\ B {\bf 544}, 357 (1999) [arXiv:hep-lat/9809004].



\bibitem{Manton:1979kb}
N.~S.~Manton,
Nucl.\ Phys.\ B {\bf 158}, 141 (1979).

\bibitem{Hosotani:1988bm}
Y.~Hosotani,
Annals Phys.\  {\bf 190}, 233 (1989).

\bibitem{Li:2002iw}
L.~F.~Li,
arXiv:hep-ph/0210063.


\bibitem{Brown:kk}
G.~E.~Brown and M.~Rho,
Phys.\ Rev.\ Lett.\  {\bf 66}, 2720 (1991). T.~Hatsuda and
S.H.~Lee, Phys.\ Rev.\ {bf C46} (1992) 34.

\bibitem{Langfeld} K. Langfeld, H. Reinhardt and M. Rho, Nucl.
Phys. {\bf A662} (1997) 620; K. Langfeld, Nucl. Phys. {\bf A642}
96c.

\bibitem{Harada:2000kb}
M.~Harada and K.~Yamawaki,
Phys.\ Rev.\ Lett.\  {\bf 86}, 757 (2001) [arXiv:hep-ph/0010207].





\bibitem{BKY}  M.~Bando, T.~Kugo and K.~Yamawaki, Phys.~Rept. {\bf 164}, 217
(1988).


\bibitem{Sannino:2002wp}
F.~Sannino,
arXiv:hep-ph/0211367. To appear in Phys.~Rev.~D.



\bibitem{Sannino:2001fd}
F.~Sannino and W.~Schafer,
Phys.\ Lett.\ B {\bf 527}, 142 (2002) [arXiv:hep-ph/0111098].


\bibitem{Lenaghan:2001sd}
J.~T.~Lenaghan, F.~Sannino and K.~Splittorff,
Phys.\ Rev.\ D {\bf 65}, 054002 (2002) [arXiv:hep-ph/0107099].







\bibitem{Alles:2002st}
B.~Alles, M.~D'Elia, M.~P.~Lombardo and M.~Pepe,
arXiv:hep-lat/0210039. (See references therein for 2 color QCD at
high baryon chemical potential)


\bibitem{Muroya:2002ry}
S.~Muroya, A.~Nakamura and C.~Nonaka,
arXiv:hep-lat/0211010.



\bibitem{ARS}  T.~Appelquist, P.S.~Rodrigues da Silva and F.~Sannino,
Phys.~Rev.~D{\bf 60}, 116007 (1999), {\tt hep-ph/9906555}.


\bibitem{DRS} Z.~Duan, P.S.~Rodrigues da Silva and F.~Sannino, Nucl.~Phys.~{\bf B} 592, 371
(2001), {\tt hep-ph/0001303}.


\bibitem{Schafer:2001bq}
T.~Schafer, D.~T.~Son, M.~A.~Stephanov, D.~Toublan and
J.~J.~Verbaarschot,
{\tt hep-ph/0108210}.
V.~A.~Miransky and I.~A.~Shovkovy,
{\tt hep-ph/0108178}.


\bibitem{Nielsen}
H.B.~Nielsen and S.~Chadha,
Nucl.\ Phys.\ {\bf B105}, 445 (1976).


\bibitem{kapusta}
J.~I.~Kapusta, {\rm Finite-temperature field theory}, Cambridge
Monographs on Mathematical Physics (1993).






\bibitem{Splittorff:2002xn}
K.~Splittorff, D.~Toublan and J.~J.~Verbaarschot,
Nucl.\ Phys.\ B {\bf 639}, 524 (2002) [arXiv:hep-ph/0204076]. See
also references therein.


\bibitem{Hands:2003dh}
S.~Hands, J.~B.~Kogut, C.~G.~Strouthos and T.~N.~Tran,
arXiv:hep-lat/0302021.


\bibitem{Moffat:2002nu}
J.~W.~Moffat,
arXiv:hep-th/0211167.



\end{thebibliography}
\end{document}